\documentclass[fleqn,10pt]{SelfArx} 

\usepackage[english]{babel} 

\usepackage{lipsum} 

\definecolor{color1}{RGB}{0,0,90} 
\definecolor{color2}{RGB}{0,20,20} 

\newcommand\aj{\ref@jnl{AJ}}
\newcommand\araa{\ref@jnl{ARA\&A}}
\newcommand\apj{\ref@jnl{ApJ}}
\newcommand\apjl{\ref@jnl{ApJL}}     
\newcommand\apjs{\ref@jnl{ApJS}}
\newcommand\ao{\ref@jnl{ApOpt}}
\newcommand\apss{\ref@jnl{Ap\&SS}}
\newcommand\aap{\ref@jnl{A\&A}}
\newcommand\aapr{\ref@jnl{A\&A~Rv}}
\newcommand\aaps{\ref@jnl{A\&AS}}
\newcommand\azh{\ref@jnl{AZh}}
\newcommand\baas{\ref@jnl{BAAS}}
\newcommand\icarus{\ref@jnl{Icarus}}
\newcommand\jaavso{\ref@jnl{JAAVSO}}  
\newcommand\jrasc{\ref@jnl{JRASC}}
\newcommand\memras{\ref@jnl{MmRAS}}
\newcommand\mnras{\ref@jnl{MNRAS}}
\newcommand\pra{\ref@jnl{PhRvA}}
\newcommand\prb{\ref@jnl{PhRvB}}
\newcommand\prc{\ref@jnl{PhRvC}}
\newcommand\prd{\ref@jnl{PhRvD}}
\newcommand\pre{\ref@jnl{PhRvE}}
\newcommand\prl{\ref@jnl{PhRvL}}
\newcommand\pasp{\ref@jnl{PASP}}
\newcommand\pasj{\ref@jnl{PASJ}}
\newcommand\qjras{\ref@jnl{QJRAS}}
\newcommand\skytel{\ref@jnl{S\&T}}
\newcommand\solphys{\ref@jnl{SoPh}}
\newcommand\sovast{\ref@jnl{Soviet~Ast.}}
\newcommand\ssr{\ref@jnl{SSRv}}
\newcommand\zap{\ref@jnl{ZA}}
\newcommand\nat{\ref@jnl{Nature}}
\newcommand\iaucirc{\ref@jnl{IAUC}}
\newcommand\aplett{\ref@jnl{Astrophys.~Lett.}}
\newcommand\apspr{\ref@jnl{Astrophys.~Space~Phys.~Res.}}
\newcommand\bain{\ref@jnl{BAN}}
\newcommand\fcp{\ref@jnl{FCPh}}
\newcommand\gca{\ref@jnl{GeoCoA}}
\newcommand\grl{\ref@jnl{Geophys.~Res.~Lett.}}
\newcommand\jcp{\ref@jnl{JChPh}}
\newcommand\jgr{\ref@jnl{J.~Geophys.~Res.}}
\newcommand\jqsrt{\ref@jnl{JQSRT}}
\newcommand\memsai{\ref@jnl{MmSAI}}
\newcommand\nphysa{\ref@jnl{NuPhA}}
\newcommand\physrep{\ref@jnl{PhR}}
\newcommand\physscr{\ref@jnl{PhyS}}
\newcommand\planss{\ref@jnl{Planet.~Space~Sci.}}
\newcommand\procspie{\ref@jnl{Proc.~SPIE}}

\newcommand\actaa{\ref@jnl{AcA}}
\newcommand\caa{\ref@jnl{ChA\&A}}
\newcommand\cjaa{\ref@jnl{ChJA\&A}}
\newcommand\jcap{\ref@jnl{JCAP}}
\newcommand\na{\ref@jnl{NewA}}
\newcommand\nar{\ref@jnl{NewAR}}
\newcommand\pasa{\ref@jnl{PASA}}
\newcommand\rmxaa{\ref@jnl{RMxAA}}

\newcommand\maps{\ref@jnl{M\&PS}}
\newcommand\aas{\ref@jnl{AAS Meeting Abstracts}}
\newcommand\dps{\ref@jnl{AAS/DPS Meeting Abstracts}}

\usepackage{hyperref} 

\hypersetup{
	hidelinks,
	colorlinks,
	breaklinks=true,
	urlcolor=color2,
	citecolor=color1,
	linkcolor=color1,
	bookmarksopen=false,
	pdftitle={Title},
	pdfauthor={Author},
}

\JournalInfo{INAF Technical Reports series, num. 40, 2020} 
\Archive{https://openaccess.inaf.it/handle/20.500.12386/27715} 

\PaperTitle{SWELTO - Space WEather Laboratory \\
	in Turin Observatory} 

\Authors{A. Bemporad\textsuperscript{1}*, 
	L. Abbo\textsuperscript{1}, 
	D. Barghini\textsuperscript{2},
	C. Benna\textsuperscript{1}, 
	R. Biondo\textsuperscript{3},
	D. Bonino\textsuperscript{1}, 
	G. Capobianco\textsuperscript{1}, 
	F. Carella\textsuperscript{1},
	A. Cora\textsuperscript{1}, 
	S. Fineschi\textsuperscript{1},
	F. Frassati\textsuperscript{1}, 
	D. Gardiol\textsuperscript{1}, 
	S. Giordano\textsuperscript{1}, 
	A. Liberatore\textsuperscript{1}, 
	S. Mancuso\textsuperscript{1},
	A. Mignone\textsuperscript{1}, 
	S. Rasetti\textsuperscript{1},
	F. Reale\textsuperscript{3},
	A. Riva\textsuperscript{1}, 
	F. Salvati\textsuperscript{1}, 
	R. Susino\textsuperscript{1}, 
	A. Volpicelli\textsuperscript{1}, 
	L. Zangrilli\textsuperscript{1}
} 

\affiliation{\textsuperscript{1}\textit{INAF-Turin Astrophysical Observatory, via Osservatorio 20, 10025, Pino Torinese (TO), Italy}} 
\affiliation{\textsuperscript{2}\textit{Turin University - Physics Department, via Pietro Giuria 1, 10125 Turin, Italy}} 
\affiliation{\textsuperscript{3}\textit{Palermo University - Physics \& Chemistry Department, Piazza del Parlamento 1, 90134 Palermo, Italy}} 
\affiliation{*\textbf{Corresponding author}: alessandro.bemporad@inaf.it} 

\newcommand{\keywordname}{Keywords} 

\Abstract{SWELTO - \textit{Space WEather Laboratory in Turin Observatory} is a conceptual framework where new ideas for the analysis of space-based and ground-based data are developed and tested. The input data are (but not limited to) remote sensing observations (EUV images of the solar disk, Visible Light coronagraphic images, radio dynamic spectra, etc...), in situ plasma measurements (interplanetary plasma density, velocity, magnetic field, etc...), as well as measurements acquired by local sensors and detectors (radio antenna, fluxgate magnetometer, full-sky cameras, located in OATo). The output products are automatic identification, tracking, and monitoring of solar stationary and dynamic features near the Sun (coronal holes, active regions, coronal mass ejections, etc...), and in the interplanetary medium (shocks, plasmoids, corotating interaction regions, etc...), as well as reconstructions of the interplanetary medium where solar disturbances may propagate from the Sun to the Earth and beyond. These are based both on empirical models and numerical MHD simulations. The aim of SWELTO is not only to test new data analysis methods for future application for Space Weather monitoring and prediction purposes, but also to procure, test and deploy new ground-based instrumentation to monitor the ionospheric and geomagnetic responses to solar activity. Moreover, people involved in SWELTO are active in outreach to disseminate the topics related with Space Weather to students and the general public.}

\begin{document}

\maketitle 

\tableofcontents 

\thispagestyle{empty} 


\section{Introduction}
\label{sec:overview}

A huge amount of data are currently available through the online archives and databases, acquired by various observatories both ground- and space-based. These data are currently distributed with a delay (or latency) from the acquisition time as low as just a few hours, therefore in near-realtime. However, our ability to reconstruct the current physical conditions of the interplanetary medium, to predict solar flares, Coronal Mass Ejections (CMEs), streams of Solar Energetic Particles (SEPs), their propagation times from the Sun to Earth, and their effects on the Earth are still quite limited \cite{riley2018}. At national level, the scientific community in Italy has been very active recently in these fields, and there are already many local infrastructures providing ground-based data, realtime data analysis and modelling, and now all these efforts are being coordinated by the Italian Space Agency (ASI) to develop a national scientific Space Weather data centre to encourage synergies between different science teams \cite{plainaki2020}. 

The solar physics group of the INAF Turin Observatory (OATo) has a consolidated experience in long-term data management, for instance with the archive \href{http://solar.oato.inaf.it/}{SOLAR} - SOHO Long term Archive \cite{cora2003}, which also includes the \href{http://solar.oato.inaf.it/UVCS_CME/index.html}{UVCS CME catalog} \cite{giordano2013}, and participation in the SOLARNET archive \cite{volpicelli2006}. The group has also experience in the development of tools and algorithms for the automated data analysis and diagnostic, such as the \href{https://www.cfa.harvard.edu/uvcs/get_involved/DAS40.html}{Data Analysis Software} for the SOHO/UVCS instrument, and for the Heliospheric Space Weather Center \cite{bemporad2018}, and more recently the on-board algorithms for Solar Orbiter/Metis \cite{bemporad2014} and for PROBA-3 metrological systems \cite{casti2019}, . 

Moreover, the group in Turin has a long term experience in the development and testing of new instrumentation and calibration facilities, such as Solar Orbiter/Metis coronagraph \cite{fineschi2020}, the OPSys calibration facility \cite{capobianco2019}, calibration of the metrological systems for PROBA-3 \cite{capobianco2016, loreggia2016} and others. All this experience is also supported by fundamental research carried out by the group over the last decades in particular in the origin and early evolution of the main driving solar phenomena related with Space Weather effects: the fast and slow solar wind flows \cite{zangrilli2016, abbo2016}, CMEs \cite{bemporad2007} and CME-driven shocks \cite{mancuso2002, susino2015}.

\begin{figure}[!t]
	\begin{center}
		\includegraphics[width=0.4\textwidth]{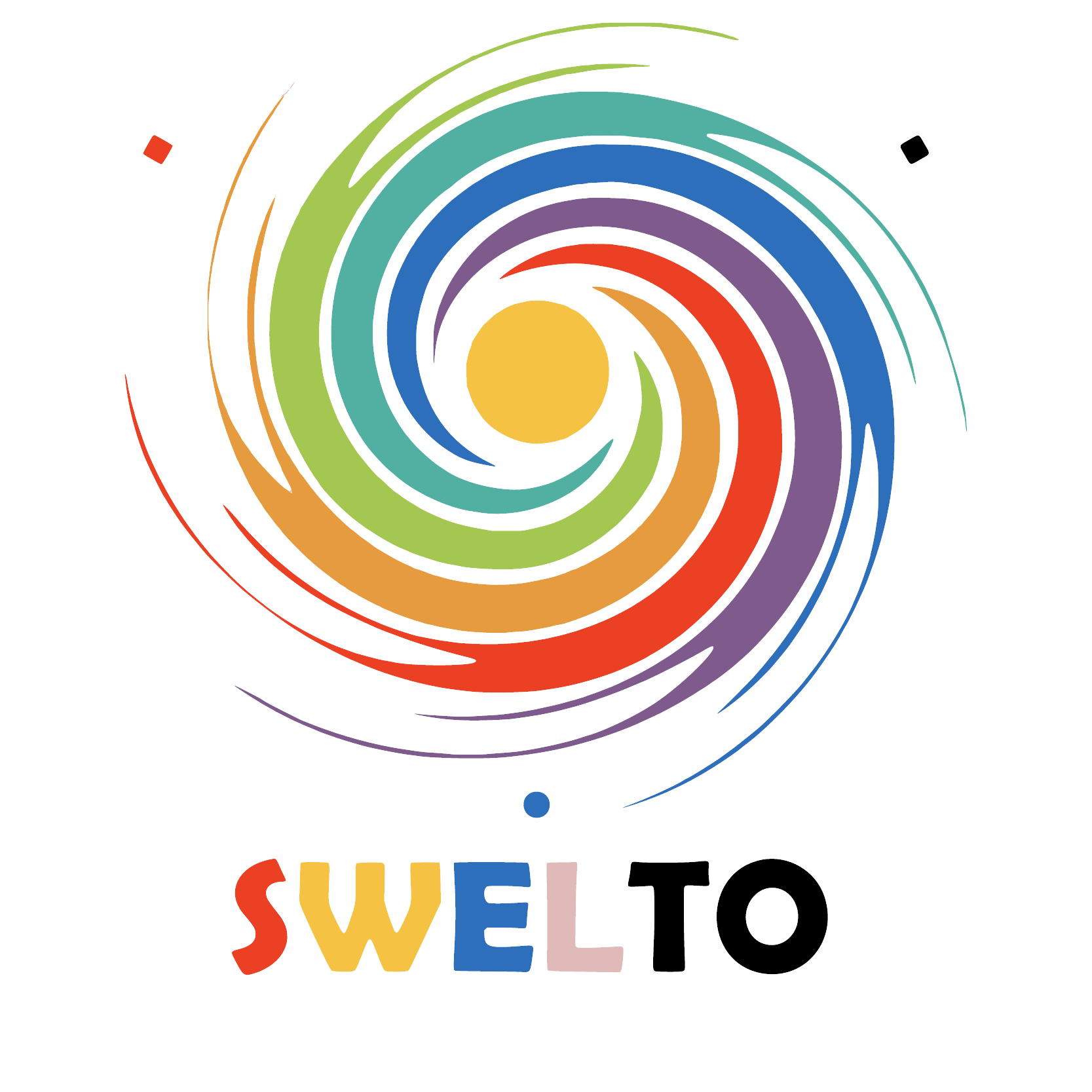} 
	\end{center}
	\caption{The SWELTO logo.}
	\label{fig:logo}
\end{figure}
Now, all this experience is being coordinated in a common effort aimed at developing new tools and routines for the automated data analysis for possible future Space Weather applications under the SWELTO project (Space WEather Laboratory in Turin Observatory). In particular, the SWELTO project has four main purposes: 
\begin{itemize}
	\item to develop, test, and implement new routines for the automated analysis of data acquired in quasi-realtime from space to provide information relative to the condition on the Sun and interplanetary medium useful to predict phenomena of interest for Space Weather events in space;
	\item to calibrate, test, and deploy new sensors (or employ already existing sensors) acquiring measurements to characterize possible geomagnetic, ionospheric, and atmospheric disturbances, hence effects related with Space Weather events in the nearby Earth environment;
	\item to involve other already existing projects and assets looking for possible new applications not considered so far of interest for Space Weather purposes;
	\item to promote outreach activities to involve students and the general public in Space Weather topics.
\end{itemize}
The above activities are aimed not only for future possible quasi-realtime Space Weather applications, but also to perform statistical analyses on the available archived data covering (with SOHO and SDO missions) more than two solar activity cycles. Moreover, the data provided by the local sensors can be potentially employed for outreach projects with high-school students, thus allowing also dissemination of solar physics, heliophysics and Space Weather science.

This article summarizes the actual status of the SWELTO project, the routines developed so far, those still under development, the sensors actually under calibration, and future developments. Each Section in this article provides as sub-title the names of the SWELTO Team members that have been directly involved with the described activities.
\begin{figure*}[!t]
	\centering
	\includegraphics[width=\linewidth]{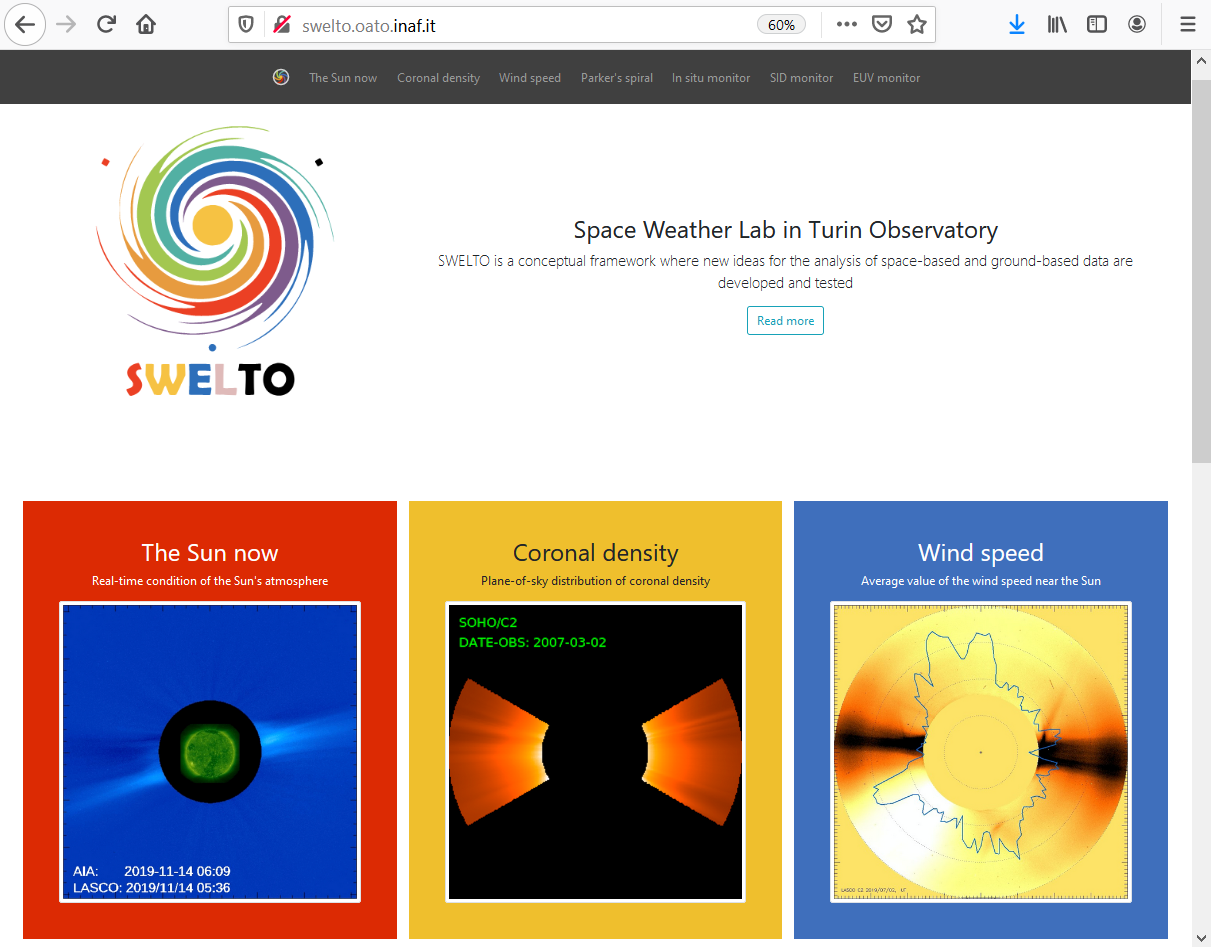} 
	\caption{A screenshot showing the \href{http://swelto.oato.inaf.it/}{SWELTO} portal.}
	\label{fig:portal}
\end{figure*}

\section{SWELTO portal and logo}

\subsection{SWELTO logo}
\textit{(Bemporad A., Susino R.)}

The SWELTO logo (Figure \ref{fig:logo}) has been conceived to represent what are the main challenges faced by the Project. The logo represents first of all the Sun (obviously the yellow disk) surrounded by a spiral representing ideally the Parker's spiral of interplanetary magnetic field. The spiral is represented with multiple different colors, to mimic the fact that the SWELTO project consists of multiple tasks (called \textit{modules}), where in each task a different physical problem (e.g. to measure coronal densities, velocities, identify solar eruptions, etc...) is faced and tentatively solved with a specific routine or automatic data analysis pipeline. Moreover, the spiral is surrounded by a filled circle at the bottom (representing any possible astronomical object orbiting the Sun and immersed in the interplanetary medium) and two small boxes at the top (representing any possible spacecraft orbiting the Sun and dedicated to the study of the Sun and the Heliosphere).

\subsection{SWELTO portal}
\textit{(Susino R.)}

The SWELTO portal (Figure \ref{fig:portal}) has been designed to provide a quick access to the products coming out from the project. From the portal it is possible to visualize the latest outputs from each module of the SWELTO project (different modules are represented by different colors matching those in the spiral logo). A click on the module allows to access to the specific page briefly describing how the products are derived in each module (input data and data analysis methods), physical meaning, limits, and approximations. If present, a link to the specific paper or technical note describing the technique is also provided.

The real-time results from the SWELTO project shown in the portal are now going to be displayed daily on a Multimedia digital stand (a 42inch multi-touch screen $1920 \times 1080$ pixels at 60Hz with integrated WiFi connection), actually under procurement. This multi-purpose stand (that will be used not only to display the real-time products from SWELTO, but also for future outreach events in our Institute) will be installed at the entrance of the office building.

\section{SWELTO automated tools}

\subsection{SWELTO workstation}
\textit{(Frassati F., Giordano S., Rasetti S., Salvati F., Volpicelli A.)}

For the SWELTO project a dedicated Workstation has been procured in May 2018. In particular, the SWELTO Workstation has a processor AMD Ryzen 7 1700 working at 3 GHz with 8-cores, 64GB of RAM (4 slots with 16GB each), a solid state drive (SSD) with 120GB of memory where the Linux Ubuntu distribution operating system is installed, and a 2TB HDD where all the downloaded data, the created pipelines, and the data products are stored. 
The Workstation has been connected on a dedicated network, to keep the machine connected 24h to internet and also guarantee that the security of the Institute's network cannot be violated passing through the SWELTO workstation. A separate account for each researcher collaborating actively on the SWELTO pipelines has been created. IDL and the \href{https://sohowww.nascom.nasa.gov/solarsoft/}{SolarSoftware} have been installed, as well as other programs such as \href{http://plutocode.ph.unito.it/}{PLUTO} to perform numerical MHD simulations, and \href{https://root.cern.ch/downloading-root}{ROOT} to analyze cosmic ray particle data.

\subsection{Structure of codes}
\textit{(Giordano S.)}

SWELTO pipelines are organized to perform automatically and continuously the near-real-time data collection and the production and display of products in a web service available to the community. 

The automated processes are scheduled by {\it{crontab}} script which performs the following steps:
\begin{enumerate}
	\item download data from near-real-time resources;
	\item select, reduce, and calibrate data;
	\item apply developed procedures and algorithms;
	\item save outputs (plots, images and movies) ready for web portal.
\end{enumerate}
The structure of the code is designed in order to allow the easy implementation of new ideas or algorithms, both for data analysis and simulations. Moreover, the graphical outputs, intended for being daily updated and on-line available, can be stored and used for scientific purposes.

\begin{figure}[!t]
	\centering
	\includegraphics[width=\linewidth]{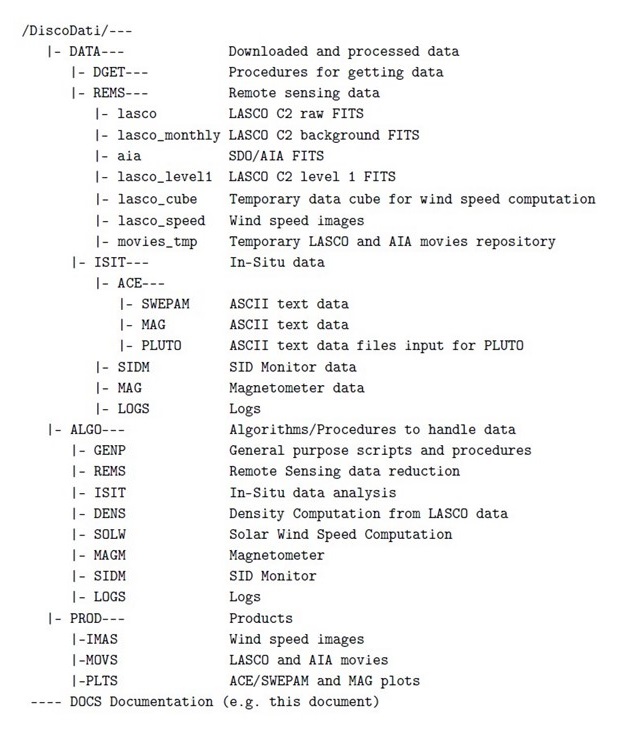} 
	\caption{The general organization of tools and data stored in the SWELTO workstation.}
	\label{fig:structure}
\end{figure}

The organization in different folders in the main HD ({\it{/DiscoDati/}}) is represented in Figure \ref{fig:structure}. The content of each folder is described into a README file in the folder itself.

The results of the pipeline called by {\it{crontab}} are daily outputs in the folder
/DiscoDati/PROD/ that are automatically copied into 
\begin{verbatim}
	/var/website/---
	|- plots
	|- images
	|- movies
\end{verbatim}
then available at the URL \href{http://swelto.oato.inaf.it/}{http://swelto.oato.inaf.it}.
\begin{figure}[!thb]
	\centering
	\includegraphics[width=0.32\linewidth]{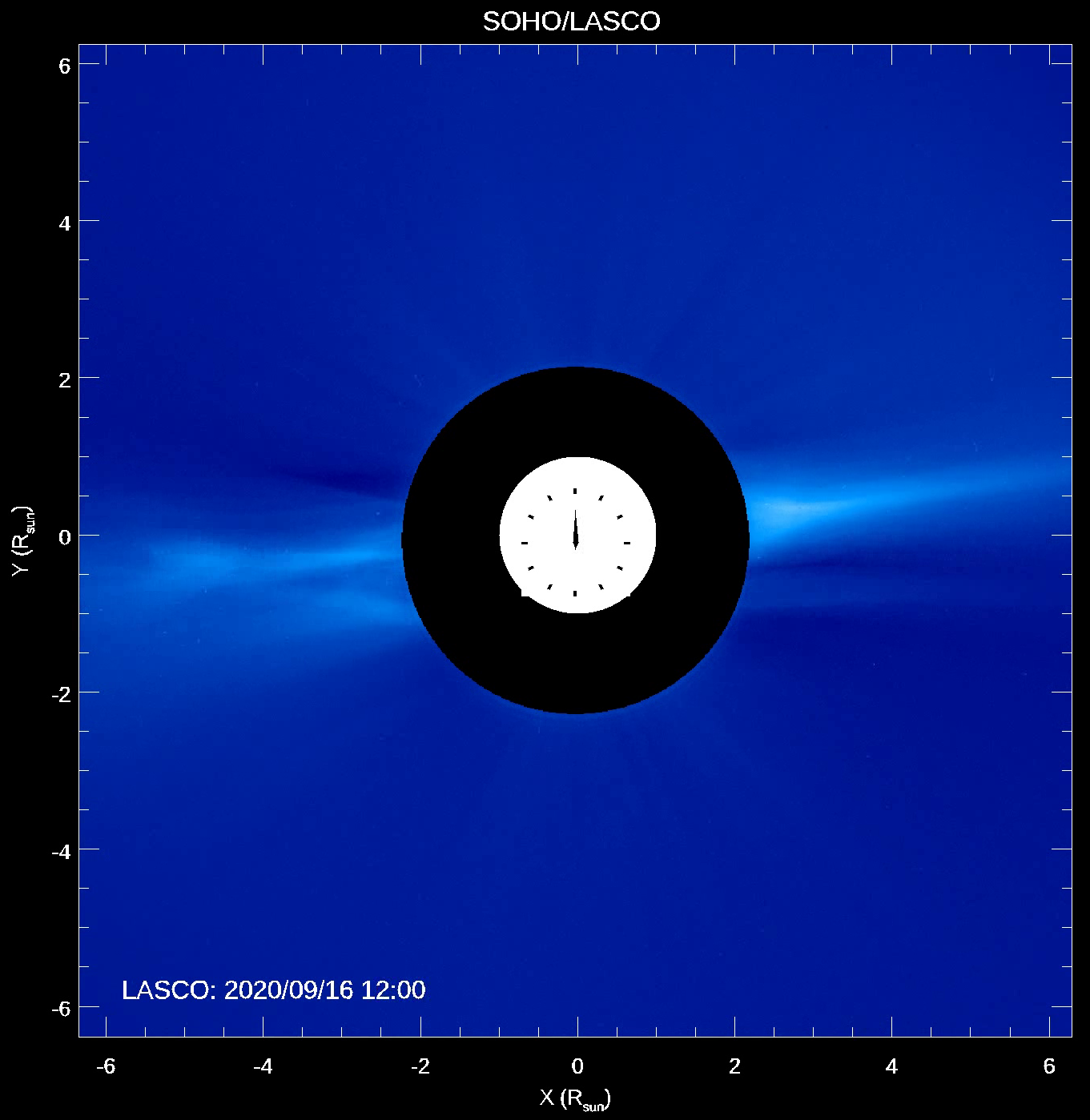} 
	\includegraphics[width=0.32\linewidth]{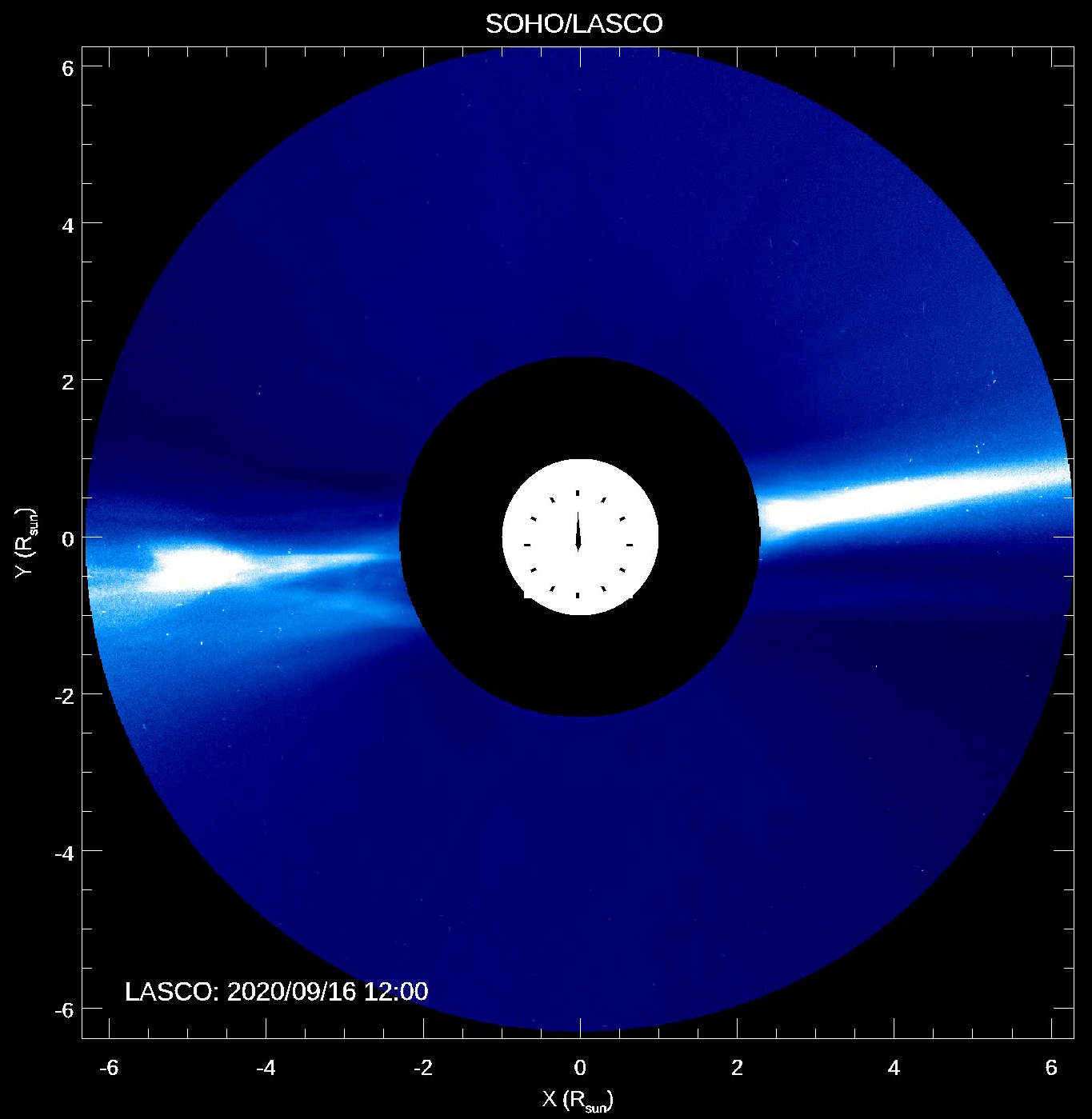} 
	\includegraphics[width=0.32\linewidth]{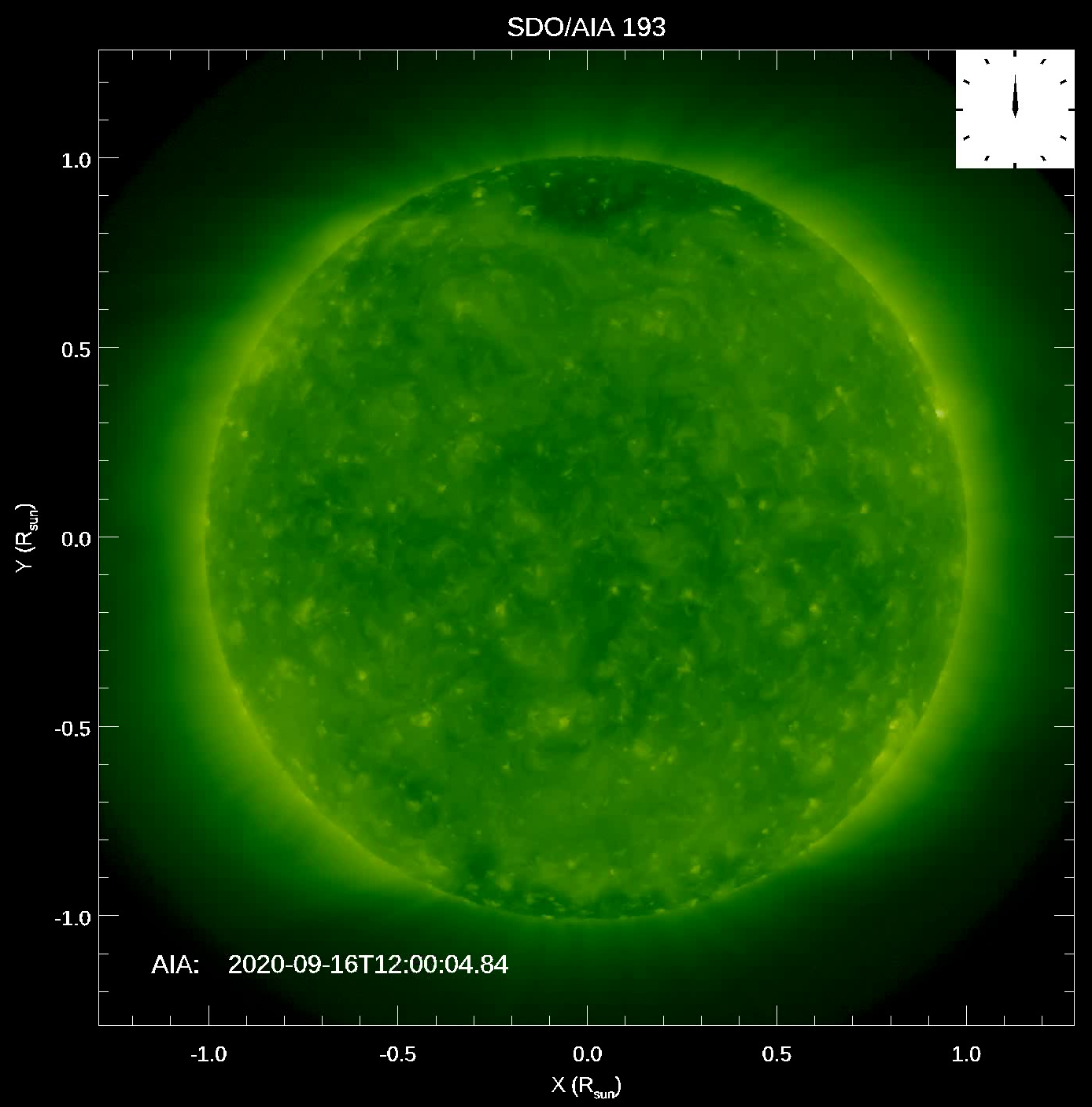} 
	\caption{SunNow module: frame from last 3 day movie, left panel: LASCO C2, central panel: LASCO C2 filtered, right panel: AIA 193}
	\label{fig:SunNow1}
\end{figure}

\subsection{SunNow module}
\textit{(Giordano S., Bemporad A.)}

This module has the purpose to provide a real-time view of current conditions on the Sun. To this end, each day the SWELTO workstation updates a local database of images acquired in particular by the SDO/AIA 193 and SOHO/LASCO-C2 instruments; only the last 3 days of data are kept on the local database. For AIA the local database is populated in particular with quicklook level 1.5 images (1024 $\times$ 1024 pixels) acquired with the 193 filter and a cadence by 3 minutes. Total brightness LASCO data are downloaded as level 0.5 files and stored. Then they are calibrated to level 1.0, and the F-corona is subtracted, on the basis of a monthly F-corona model released from LASCO team. 

This module produces the calibrated LASCO C2 data files used to calculate electron density maps (see CorDens module) and to create movies showing the most recent evolution of the solar inner and intermediate corona as seen from the Sun-Earth line direction. 
In particular, this module creates the last 3 days movies of following observables:
\begin{itemize}
	\item LASCO C2 white light brigtness;
	\item LASCO C2 white light brigtness filtered;
	\item AIA 193\AA\ intensity;
	\item Combined LASCO C2 and AIA 193\AA.
\end{itemize}
\begin{figure*}[!thb]
	\centering
	\includegraphics[width=0.9\linewidth]{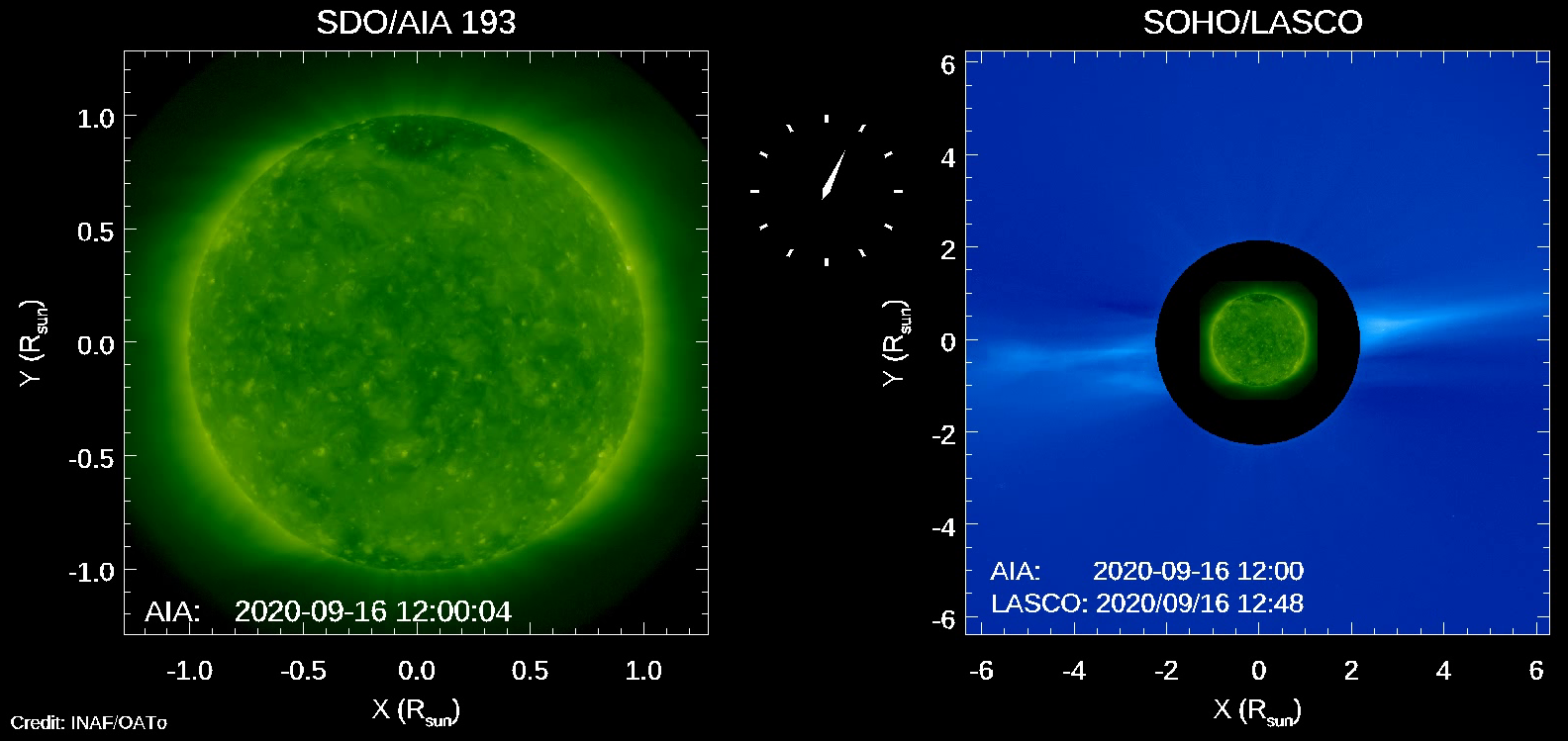} 
	\caption{SunNow module: frame from Combined LASCO and AIA last 3 day movie}
	\label{fig:SunNow2}
\end{figure*}

Figures \ref{fig:SunNow1} and \ref{fig:SunNow2} shows single frames from the movies available on-line. All these products are available to the public via the SWELTO project portal described before.

\subsection{EUVmonitor module}
\textit{(Carella F., Bemporad A.)}

\begin{figure}[b!]
	\centering
	\includegraphics[width=\linewidth]{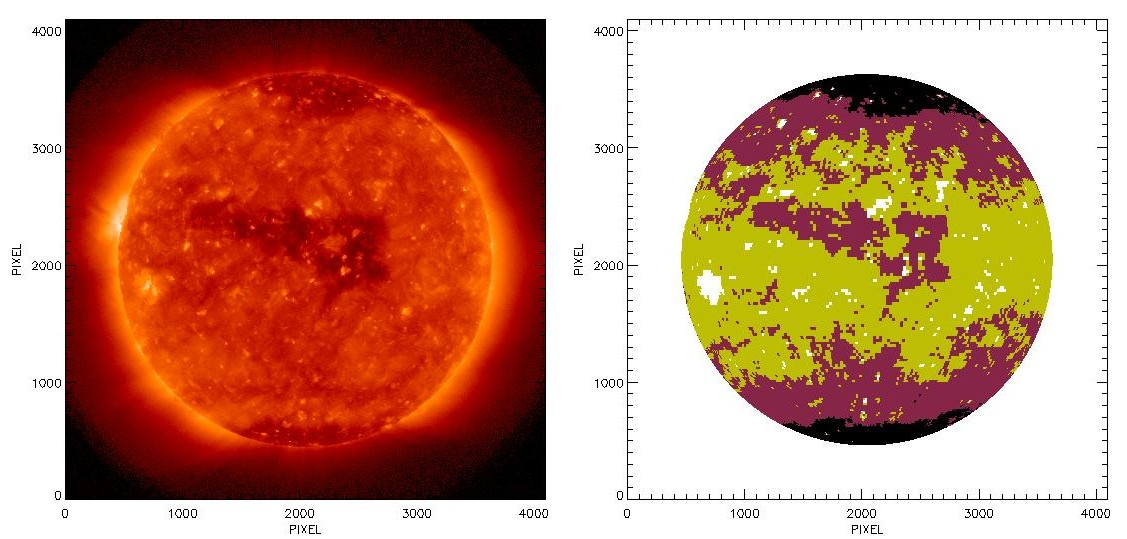}
	\caption{FCM application on AIA 193 Å minimum image (05-04-2018 18:00:00) with an equatorial coronal hole: in black PCH, in purple MCH, in green QS and in white AR.}
	\label{fig:fcmres}
\end{figure}
This module is meant for the automatic identification of structures of the Sun based on EUV images and photospheric magnetograms, utilizing a FCM (Fuzzy C-Means) algorithm. Before the algorithm's application, a pre-processing is applied to AIA 193 \AA\ images and HMI LOS magnetograms aquired by SDO, and a routine based on a geometrical approach corrects for the limb-brightening effects on EUV images.\\
The images and the magnetograms are then analyzed pixel-by-pixel by determining the degree of membership of each pixel to one of clusters, previously defined based on the analysis of a sample training dataset. After this, a segmentation is created thanks to the outputs of FCM algorithm. An example of FCM application and its results is shown in Figure \ref{fig:fcmres}. All of these routines are written in IDL programming language. More details are provided in \cite{carella2020}.

\subsection{CorDens module}
\textit{(Zangrilli L., Giordano S.)}

This module is aimed at providing a quasi real-time determination of the 2D distribution of coronal densities on the plane of the sky. The electron density distribution in the corona is obtained, with a regular cadence of one hour, from LASCO/C2 images, covering the FoV from 2 to 6 solar radii, and in coronal sectors 60 degrees wide across the equator. The aim is to provide electron density maps with the same cadence of images in the Sun-Now Module, and this can be done calculating the density from the total brightness data, and not starting from the pB measurements, which are acquired once per day. We used a standard inversion technique of the integral along the LOS of the total K-corona, which is due to Thomson scattering of the solar disk radiation from the coronal electrons. The K-coronal brightness is derived from the total coronal brightness, after subtraction of the F-corona distribution. The same procedure can be applied to LASCO/C3, in order to extend the electron density maps to larger heliocentric distances, virtually up to 30 solar radii, even if the strong dominance of F-corona at those distances makes the electron density estimates prone to larger errors.

Future developments could foresee the use of different space-born coronagraph, such as STEREO/COR instruments, and the use of ground-based coronagraph, for the inner solar corona.
\begin{figure*}[!thb]
	\centering
	\includegraphics[width=\linewidth]{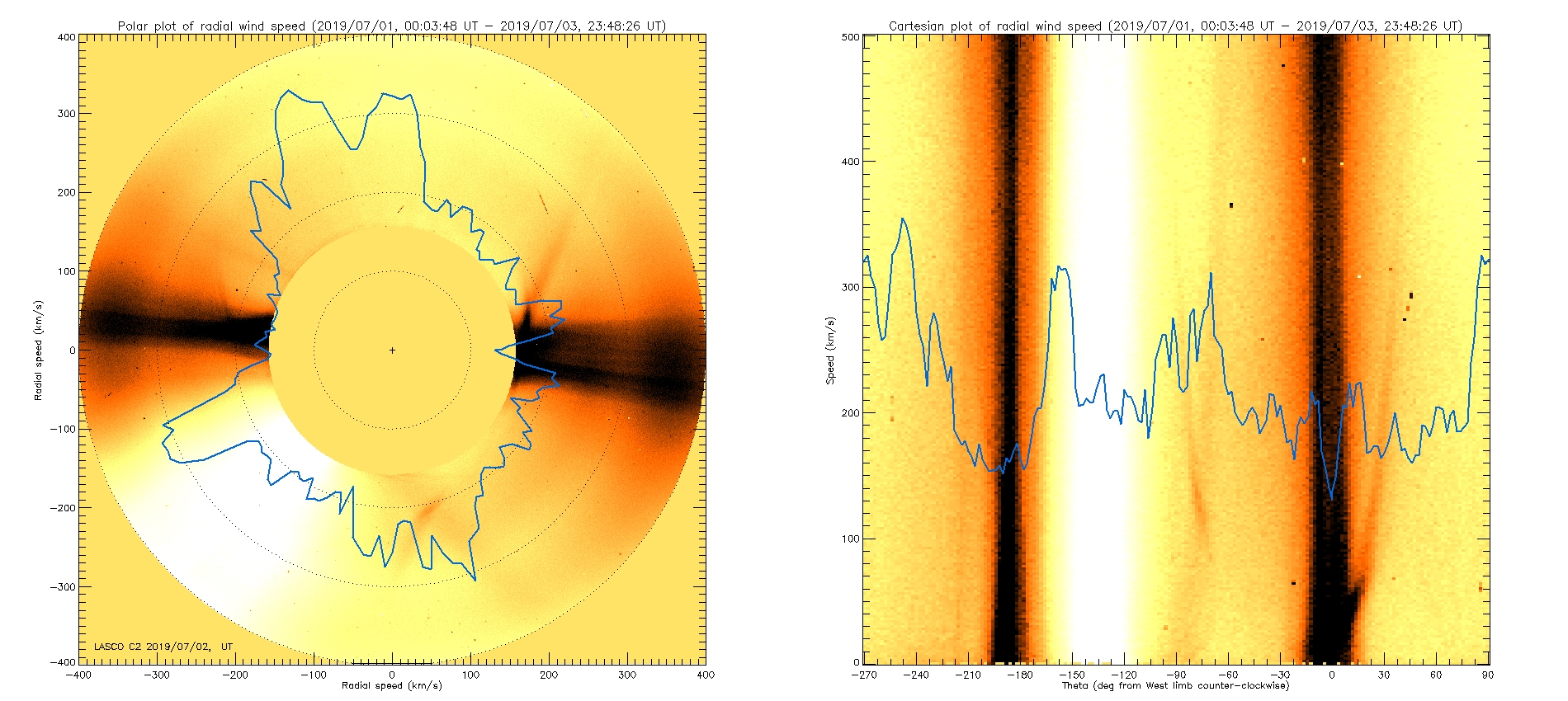} 
	\caption{Example of output products from the CorSpeed module: latitudinal distribution of solar wind speed (km/s) measured with cross-correlation and plotted in polar coordinates over a LASCO-C2 image (left) and over a polar transformation of the same image (right).}
	\label{fig:CorSpeed}
\end{figure*}

\subsection{CorSpeed module}
\textit{(Abbo L., Bemporad A.)}

The aim of this module is to derive a measurement for the projected radial speed of plasma blobs and/or density inhomogeneities in the solar corona, by assuming that these features can be employed as tracers of the local solar wind speed \cite{abbo2016}. The input data are sequences of visible light coronagraphic images, for the first test presented here we employed data acquired by the SOHO/LASCO-C2 coronagraph. All the total brightness images (orange filter) acquired over a period of 3 days are downloaded, calibrated with standard SolarSoftware routines, transformed in polar coordinates, and packed in a single datacube. The time evolution of intensities observed in each single pixel is de-trended to remove variations related with large scale coronal features, and identify small scale features such as plasma blobs and density inhomogeneities. At each single latitude, all the radial intensity profiles observed over the selected time interval are then analyzed with cross-correlation algorithm. In particular, at any fixed latitude the time evolution observed at each altitude in the instrument field of view is cross-correlated with itself, in order to derive the time shift maximizing the cross-correlation coefficient over the whole time interval. Once this time shift is measured, the corresponding average value of the plasma speed is easily determined, simply taking into account the projected extension of the pixels in C2 images.

The resulting latitudinal distribution of the wind speed is finally shown in a polar plot superposed over one of the LASCO-C2 images normalized with a radial filter (Fig.\ref{fig:CorSpeed}). This allows a direct graphical comparison between the latitudinal distribution of the measured speeds and the latitudinal location of different coronal features, such as streamers and coronal holes. The routine works with the latest available LASCO images in quasi real-time, providing daily measurement of the actual wind speed on the plane of the sky every 6 hours. In its first implementation the routine is not able to measure tangential displacements of the plasma, hence the measurements are representative only of purely radial motions of density inhomogeneities propagating with the solar wind. This measurement (which is expected to be important only in the inner corona) will be tested and included in the future. Moreover, for future developments of this module, we will also try to employ the technique applied by \cite{cho2018} and based on the analysis with Fourier filtering of images acquired by the LASCO-C3 coronagraph.
\begin{figure*}[!thb]
	\centering
	\includegraphics[width=\linewidth]{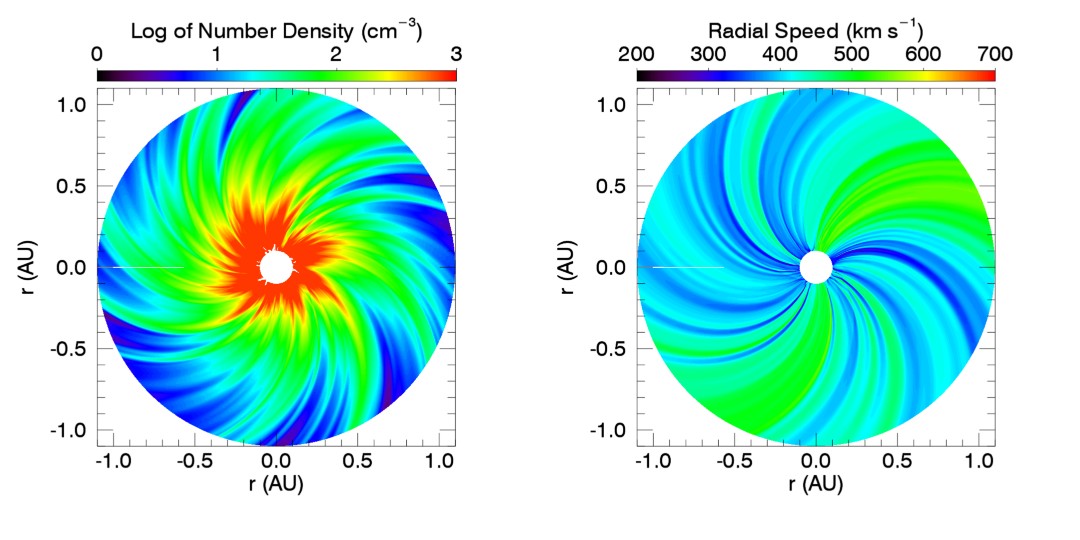}
	\caption{Ecliptic maps of plasma number density (left) and radial speed (right) simulated from 1 AU in-situ measurements collected in March 2009.}
	\label{fig:RIMAP}
\end{figure*}

\subsection{InSitu module}
\textit{(Bemporad A., Giordano S.)}

This module is aimed at providing a quasi real-time information of the actual conditions in the interplanetary space. The input data are sequences of plasma measurements acquired in quasi real-time by the ACE spacecraft located in the L1 Earth-Sun Lagrangian point. Among different acquired measurements, we selected the wind speed, density, and the three components of the magnetic field. Data acquired over the last 24 hours are extracted and plotted (1 min or 5 min average); moreover the data are used also to show actual values of the Alfv\'en speed and wind Alfv\'enic Mach number (the employment of these parameters for possible space weather forecasting applications will be tested in the near future). The module also read in input all the measurements acquired over the last 28 days, in order to cover one full rotation of the Parker spiral. These measurements are then re-sampled, and provided in output as input parameters for the “Parker spiral module”.

In the future developments this module will also provide output plots for "combined" plasma physical quantities, such as: the solar wind Alfv\'en speed, Alfv\'enic Mach number, magnetosonic Mach number, ram pressure, and the estimated stand-off distances of the Earth's bow shock and magnetopause (Chapman-Ferraro distance) along the Sun-Earth line.

\subsection{ParkerSpiral module}
\textit{(Biondo R., Reale F., Bemporad A., Mignone A.)}

The purpose of this module is to numerically reconstruct the condition of the interplanetary plasma in the Parker Spiral from 0.1 to 1.1 AU, a key step in understanding the propagation of solar disturbances. An example of output product from this module is shown in Figure \ref{fig:RIMAP}.\\
Since only very few spacecrafts measure these parameters in-situ, the determination of the plasma state in the vicinity of the Sun is the main obstacle that a reconstruction faces with. To overcome this difficulty, two main approaches have been developed: analytic ballistic back-mapping starting from in-situ measurements acquired at 1 AU (\cite{ness2001,florens2007}) and numerical forward MHD simulations starting from photospheric fields remote-sensing measurements (\cite{WSArge,WSA4,mikic2018}). Analytical methods are computationally simple and give good agreement with the large scale features reconstruction, however they need to assume stationarity, often assume the same average speed at all longitudes to avoid stream-line crossings and are not time dependent. On the other hand, numerical models offer deeper insight into the physics of solar and space-weather phenomena (\cite{macneice2018}), providing real-time evolution of the whole Spiral, but they have limited knowledge of the plasma condition at their inner boundaries.\\
Our model, RIMAP (Reverse In-situ data and MHD APproach), combines for the first time the two methods: it uses as input an analytic back-propagation of 1 AU in-situ measurements provided by the InSitu module to reconstruct the plasma parameters located at 0.1 AU, using them as the internal boundary of a MHD numerical simulation solved using the PLUTO code (\cite{PLUTO1,PLUTO2}) that propagates plasma streamlines up to 1.1 AU. 

Future developments of this module will consider the extension to the third dimension by using input plasma parameters in the inner boundary as measured from coronagraphic images with the above CorDens and CorSpeed modules. Moreover, the expansion and propagation through the simulated plasma environment of solar disturbances will be tested and integrated with a future module (to be developed) aimed at the automated identification of CMEs and determination of their main kinematical properties (speed, angular width, propagation direction) from coronagraphic images. More details are provided in \cite{biondo2020}.

\section{SWELTO sensors network}

\subsection{SIDmonitor module}
\textit{(Bemporad A., Riva A., Salvati F.)}

This module is aimed at providing real-time information about possible sudden ionospheric disturbances (SIDs) as observed with local data. A radio antenna was built, installed in the Turin Observatory, and connected with a SID monitor provided by Stanford University (in the occasion of the 2007 celebrations of International Heliophysical Year). The antenna is made with a 100m long cable rolled-up over covering a 1m$^2$ square area. The monitor band-pass is centered on the frequency of 23.4 kHz and samples the signal emitted by a strong VLF station located in Germany at Rhauderfehn (DHO station at 53$^\circ$10'N 07$^\circ$33'E). During day-time this signal propagates through multiple reflections between the ground and the D layer of ionosphere, while during night-time the D layer disappears, and the signal is reflected by the E layer located higher up, leading to a typical night-day modulation of the signal. When a solar flare occurs, the enhanced ionization due to the increased X-ray and EUV flux from the Sun results in a clear associated increase in the strength of radio signal, due to the ionospheric disturbance related with the solar flare. The module automatically reads and plot real-time measurements acquired with the SID monitor.

\begin{figure}[!h]
	\centering
	\includegraphics[width=\linewidth]{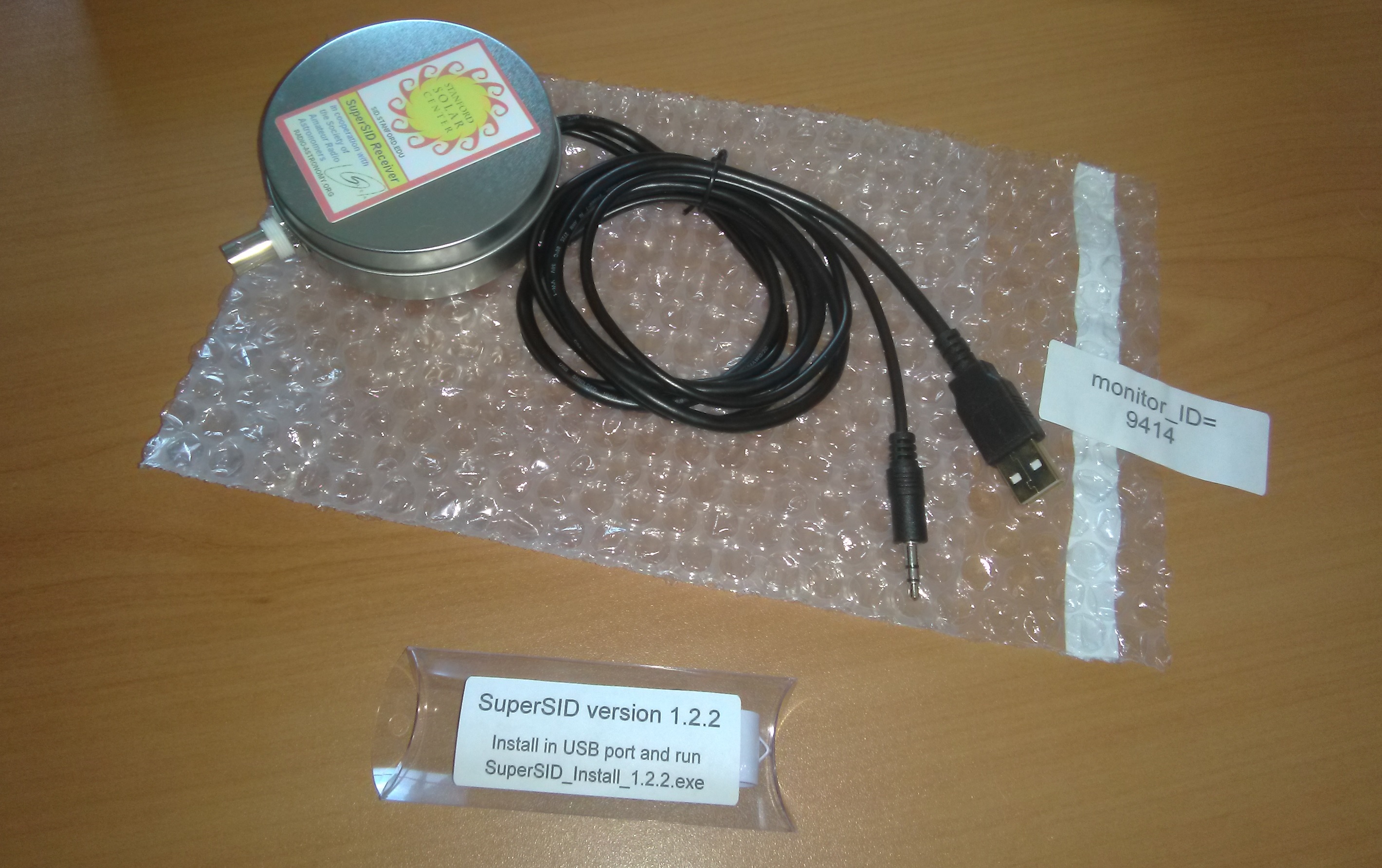} 
	\caption{The SuperSID monitor procured from  \href{http://solar-center.stanford.edu/SID/sidmonitor/}{Stanford Solar Center}.}
	\label{fig:SIDmonitor}
\end{figure}

This module is now going to be refurbished with the most-recent version of the SuperSID monitor (Fig.\ref{fig:SIDmonitor}) provided again by Stanford University as in-kind contribution to our project. This monitor will be connected with a new radio antenna (now under development) with a 200m long cable rolled-up over a hexagonal perimeter by 4.5m covering one area of about 1.46 m$^2$. This will allow to significantly improve the signal-to-noise ratio with respect to the previous antenna, as clearly pointed out by \cite{arnold2014}. The main parts of this antenna will be drawn in 3D and printed with a 3D printer available locally in our Institute.

\subsection{FGmagnetometer module}
\textit{(Capobianco G., Bemporad A.)}

The FGmagnetometer module will monitor the consequences of solar variability on Earth by measuring one of the most important markers, the geomagnetic activity. Local measurements of the geomagnetic field at mid-latitude will be performed at the  INAF-Astrophysical Observatory of Torino ($45 ^\circ 02 '29''N$, $7^\circ45'55''E$), where the expected magnitude of geomagnetic field should be 46.473 $\mu$Tesla (0.46473 Gauss), as provided by the \href{https://www.ngdc.noaa.gov/geomag/calculators/magcalc.shtml}{NOAA Geomagnetic Field Calculator}. Starting from this average reference value, the magnetometer should be able to measure fluctuations with amplitude at least on the order of 0.1 nT, hence a minimal accuracy on the order of $0.1 \times 10^{-9} / 46.473 \times 10^{-6} = 2 \times 10^{-6}$ (corresponding to 2 ppm) is required. These measurements have to be acquired with a minimal sampling rate by 10 Hz.
\begin{figure}[!thb]
	\centering
	\includegraphics[width=\linewidth]{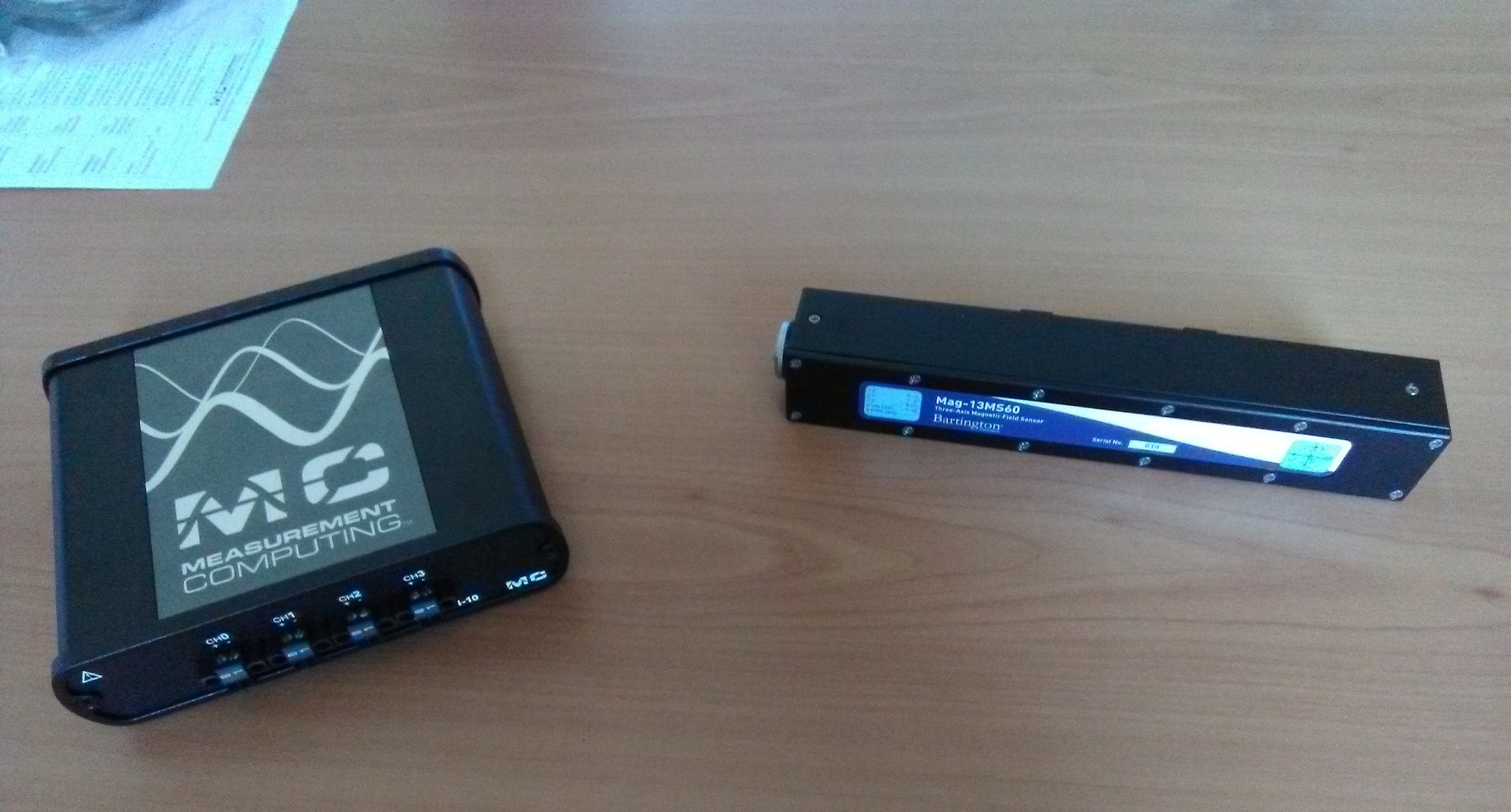} 
	\caption{The 3-axis Flux-Gate Magnetometer (right) procured from  \href{https://www.bartington.com/}{Bartington Instruments LTD} (model MAG-13MS60), and the 4-channels ADC controller (left) procured from \href{https://www.mccdaq.com/}{Measurement Computing} (model USB-2404) for the measurement of local geomagnetic disturbances.}
	\label{fig:FGmagnetometer}
\end{figure}

Given the above requirements, the magnetometer that we selected is a 3-axis Flux Gate sensor manufactured by the \href{https://www.bartington.com/}{Bartington Instruments LTD}. The signals (the 3 components of the magnetic field vector and the temperature of the magnetometer) are acquired and digitized by a 24-bit ADC procured from \href{https://www.mccdaq.com/}{Measurement Computing}. Figure \ref{fig:FGmagnetometer} shows the procured magnetometer (right) and ADC (left). The acquisition system is based on the Raspberry Pi computer. The system is compatible, except for the power and data connection, for the installation on a remote station. The software is programmable for acquisition at frequency up to 50 Ksamples/s and averages returned with the desired cadence. The current baseline is the signal acquisition at frequency of 10 Hz, with averages every 60 seconds, 3 hours and 12 hours. 

The resolution of the magnetometer is of 7 pT, when the accuracy is expected to be of about 10 pT after the temperature calibration (a residual is expected due to the accuracy in the N-S alignment of the magnetometer). The system is expected to be installed before the end of the 2020 and debugged for the spring of 2021.

\subsection{SkyMonitor module}
\textit{(Mancuso S., Gardiol D., Barghini D., Bonino D., Bemporad A.)}

In this module we explore the possible connection between energetic transient events on the Sun and a peculiar class transient events, known as sprites, detected in the Earth’s atmosphere.

Transient luminous events (TLEs) are large-scale optical events occurring in the upper-atmosphere from the top of thunderclouds up to the ionosphere.
These electrodynamic processes can be commonly divided into two major groups: lightning-induced events (sprites, elves, sprite halos) and upward discharge between a thundercloud and the lower ionosphere (blue starters, blue jets, gigantic jets).
From a morphological point of view, sprites are usually described as a group of visible columns, sometimes with branching features that make them appear jelly-fish or carrot-like.
These particular TLEs are produced by the residual quasi-static electric field following an exceptionally strong cloud-to-ground (CG) lightning (\cite{pasko1997}) and since their serendipitous discovery in 1989 (\cite{franz1990}), they have been intensively studied by a number of ground-based optical and electromagnetic observations since the discovery. 
These observational studies clarified that sprites manifest a predominately-red luminous structure that occur at mesospheric altitudes between 40 and 90 km (\cite{sentman1993}) with optical emissions that spans a time interval from a few milliseconds to a few tens of milliseconds.
Theoretical studies on the sprite generation mechanisms revealed that sprite optical emissions are produced by the excitation/relaxation processes of neutral molecules and ions (mainly N$_2$ and N$_2^+$) in the atmosphere (\cite{pasko1997}; \cite{ebert2006}).
These excitation processes are induced by collisions between these particles and the ambient electrons accelerated by the quasi-electrostatic field generated by cloud-to-ground discharges.

The atmosphere of the Earth is subject to interactions with highly energetic particles continually coming from outside the solar system and sporadically generated from the Sun. 
Solar energetic particles (SEPs) are accelerated by magnetic reconnection or stochastic mechanisms in the parent flare region or by CME-driven shocks in the solar corona and interplanetary space. 
They consists of electrons, protons and heavier ions with kinetic energies ranging from tens of keV to about 10 GeV. 
The elemental composition of SEPs changes from one event to another, but it is generally dominated (> 90 \%) by protons. 
Although these particles are known to generate ionization along their path, the subsequent physical and chemical effects in the Earth’s atmosphere are still under active investigation. 
This solar-terrestrial coupling may have implications for a number of atmospheric processes. 
For example, increased fluxes of solar energetic charged particles have been found to cause significant conductivity increases either in the thunderclouds or in the surrounding air (\cite{nicoll2014}) or in areas above thunderstorms (\cite{tacza2018}). 
In addition to atmospheric electric circuit effects, sufficiently energetic charged particles may be able to provide narrow ionization channels that trigger cloud-to-ionosphere discharges (see discussion in \cite{owens2014}), a region where TLEs are produced.

\subsection{CosmicRays module}
\textit{(Liberatore A., Bemporad A.)}


The study of cosmic rays and particles coming from outer space and from the Sun is a relevant part of the space weather field. This module has the purpose to study this cosmic radiation. It is known, in particular, how solar activity affects the Van Allen belts shape and particles density by disturbing the Earth geomagnetic field. In addiction, after the interaction of these cosmic rays with the upper Earth atmosphere layer, secondary cosmic ray shower can be produced (detectable on the ground at high altitude, and added to the standard flux of cosmic rays).

One of the main goal of the studies of this module is to evaluate the cosmic ray environmental dose variation in different part of the world and at different altitude (with particular attention to the polar regions because, due to the geomagnetic field shape, are the most exposed to space radiation~\cite{ZANINI2017149}). To date, literature about this kind of measurements -high altitude and latitude- is very poor. In particular, just few environmental dose measurements in high southern latitudes were performed. Thus, measurements at different locations and altitudes in the southern hemisphere [\textit{e.g.} Ushuaia (Argentina, 54\textdegree S, 68\textdegree W; sea level), Famatina Mountain (Argentina, $\sim 5000$ m a.s.l.), Marambio (Antarctica, 64\textdegree S, 56\textdegree W; sea level) and Dome-C (Antarctica, 64\textdegree S, 123\textdegree E; 3233 m a.s.l.)], can be useful to have a general view of the environmental dose due to cosmic radiation at the high latitude and different altitude on our planet. In particular, at Plateau Rosa (Italy; 45\textdegree N, 7\textdegree E; 3450 m a.s.l.], different dosimeters and particle detector instruments are in \textit{Testa Grigia} laboratory (when not used at others latitudes), such as an INAF modular neutron monitor~\cite{Signoretti2011} for the detection of primary cosmic rays, and the Liulin LET Spectrometer~\cite{ZANINI2019105993, DACHEV20091441}. Active instruments like a Neutron rem counter Atomtex BDKN-03 or gamma ray detectors can be also set into the laboratory for instrument test or data acquisition. Even passive detectors (e.g. a set of bubbles dosimeters) can be used. Data acquisition in Turin Observatory is possible with these active and passive detectors as well.

Furthermore, it is also known that the SAA -\textit{South Atlantic Anomaly}- area, the region in which the Van Allen belts come closest to the Earth surface, is gradually expanding~\cite{DESANTIS2012129, Hamilton2012} and, at the same time, the global geomagnetic field is decreasing significantly (particularly in the Antarctic region/high latitudes)~\cite{Lepidi2003}. These facts, combined with the decrease in solar activity found in the last solar cycles~\cite{Singh2019}, leads to an increase in the flow of cosmic rays reaching the Earth. Nevertheless, a dosimetric monitoring in the SAA area is totally absent.
For these reasons it could be performed a first study in association with solar storms and Earth magnetic field variation as well. In summary, the main scientific objectives are:
\begin{itemize}
	\item to understand the contribution of the altitude, latitude and geomagnetic field and solar activity to the cosmic ray variation and respective dose on ground (with the opportunity of a more systematic study at Plateau Rosa laboratory - 3450 m a.s.l.); 
	\item to increase the dosimetric data statistic of the Antarctic and sub-Antarctic regions;
	\item to contribute to the dosimetric monitoring of the SAA region and to understand its connection with the cosmic ray radiation measured on ground;
\end{itemize}

Particular attention is paid to the neutron component. In fact, the neutron component, due to the high "radiation weighting factors", is responsible for most of the radiological risk at the quotas of interest to the human being. Moreover, a posteriori analysis with data obtained from different solar spacecraft (Solar Orbiter, Solar Probe, etc\dots) can be performed to better  understand the Sun-Earth-ground dose connection. 

All these measures and instrumentation comes from different projects we are working on such as HALCORD (\textit{High Atitude and Latitude Cosmic Ray Dosimetry}, 2017-2019, INFN project) and CORDIAL (\textit{COsmic Ray Dosimetry In Antarctic Latitudes}, 2020-2022, PNRA project) and other in a proposal phase (\textit{e.g.} SAMADHA, \textit{South Atlantic Magnetic Anomaly Dosimetry at High Altitude}).

\section{SWELTO dissemination \& outreach}
\textit{(Benna C., Cora A., Gardiol D., Bemporad A.)}

\subsection{Dissemination and outreach activities}
Among different activities described above, the SWELTO project is also aimed at the dissemination of main topics related in general with Space Weather. This is done with the double purpose to involve students in the research being carried out with SWELTO, and also to promote the results to the general public to raise their awareness on the importance of this research. This is done with different channels, briefly listed below.

\begin{itemize}
	\item \href{https://edu.inaf.it/news/eventi/report/master-class-di-astronomia-e-astrofisica-2017/}{Astronomy and Astrophysics MasterClass}: addressed to high-school students. A group of students is hosted for one day in the facilities of the Turin University - Physics Department, and at INAF - Astrophysical Observatory of Turin. During the MasterClass the students attend astronomy and astrophysics lectures held by young scientists and researchers, and follow different activities. One of them was dedicated at the study of Coronal Mass Ejections (CMEs): students are taught to examine the UltraViolet Coronagraph Spectrometer (UVCS) FITS files and Large Angle Spectrometer Coronagraph (LASCO) images, in order to measure the speed of the CME wave front and estimating the possible arrival time on Earth. These kind of activities show very well the interaction between space and earth allowing to introduce the main space weather topics.
	\item \href{http://www.campusmfs.it/}{Campus MSF}: a Campus addressed to high-school students and entirely dedicated this year to Astronomy and Astrophysics. The Campus offers a unique opportunity for secondary school students to interface with the world of University research in an intellectually stimulating environment, where they meet and interact with University professors, astronomers and astrophysicists and with other students who share with them the same interests and passions.
\end{itemize}

The SWELTO team members are involved in teaching (for instance with a full \href{https://fisica.campusnet.unito.it/do/corsi.pl/Show?_id=a293}{University Course} dedicated to Heliophysics and Space Weather topics), and dissemination activities carried out in the Observatory (to school classes visiting our Institute or participating to the "alternanza scuola e lavoro" activities), and outside the Observatory (with stages/campuses organized with other University Departments). These activities include, in addition to the classical frontal lessons, also practical and experimental activities (such as the observation of the sun, the measurement of CME speed, and the solar constant).

Furthermore, the recent pandemic emergency has prompted staff members to develop Python scripts, to freely distribute data analysis software useful for school teachers who want to organize classrooms and to involve students even with remote activities.

\subsection{SWELTO multimedia workstation}
\begin{figure}[!thb]
	\centering
	\includegraphics[width=\linewidth]{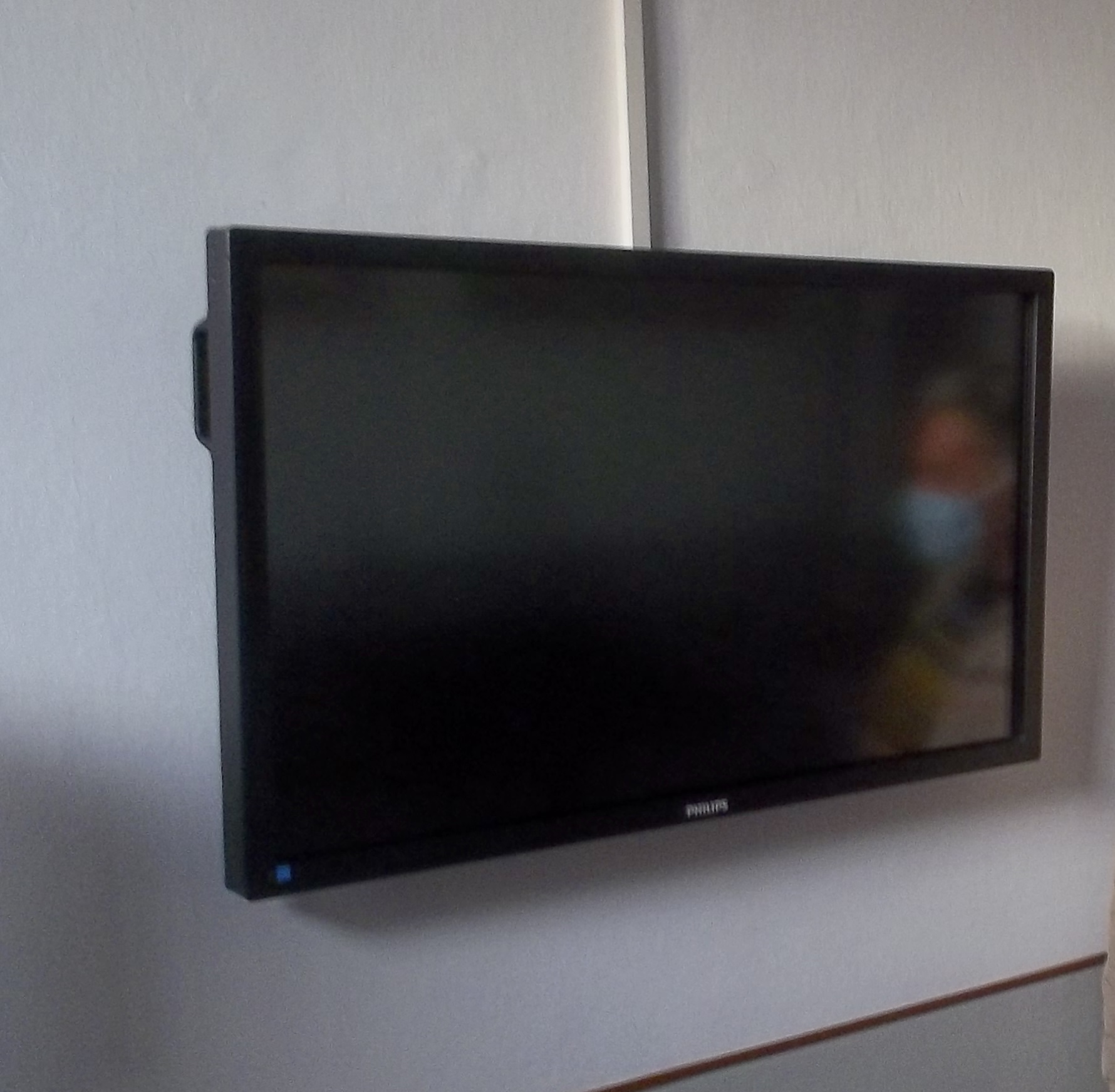} \caption{The digital signage multimedia stand procured for the dissemination of SWELTO project and visualization of real-time products.}
	\label{fig:multimedia}
\end{figure}

To help the dissemination of SWELTO project and results, a digital signage multimedia workstation has been procured (Figure \ref{fig:multimedia}). This monitor (in particular a Philips Monitor Digital Signage 43”) is equipped with a multi-touch display and network (WiFi or LAN) connectivity. Once installed in the Institute, the screen will be used to visualize in real-time the SWELTO portal and navigate through its products. The workstation will also be available for any other outreach activity that will be performed in the Institute. 

\section{Conclusions and future perspectives}
\textit{(Bemporad A., Fineschi S.)}

This technical note describes the actual status of the SWELTO project, that started at INAF-Turin Astrophysical Observatory around the end of 2017. The different modules preliminarily described here and conceived for possible future Space Weather applications, have at present different maturity levels, that are summarized below by using the following increasing categories: \textbf{Level0} TO BE CONCEIVED, \textbf{Level1} TO BE DEVELOPED, \textbf{Level2} TO BE TESTED, \textbf{Level3} UNDER ACTIVATION, \textbf{Level4} ACTIVE.

\begin{itemize}
	\item \textbf{SunNow} module status: \textbf{Level4}. The pipelines are running in real-time and the output movies are successfully produced, stored in the local disk, and freely distributed via the SWELTO portal;
	\item \textbf{EUVmonitor} module status: \textbf{Level2.5}. Further testing on the scientific reliability of results are required before the module can go to Level3. The pipelines are ready to be executed in real-time, but the download of HMI images have still to be implemented;
	\item \textbf{CorDens} module status: \textbf{Level3.5}. The pipelines are ready to be executed in real-time, and the output results have been tested, but the output density maps are not actually provided in real-time;
	\item \textbf{CorSpeed} module status: \textbf{Level2.5}. Further testing on the scientific reliability of results are required before the module can go to Level3. The pipelines are ready to be executed in real-time;
	\item \textbf{InSitu} module status: \textbf{Level4}. The pipelines are running in real-time and the output plots are successfully produced, stored in the local disk, and freely distributed via the SWELTO portal; the output data also also successfully provided as an input to the ParkerSpiral module;
	\item \textbf{ParkerSpiral} module status: \textbf{Level4}. The pipelines are running in real-time and the output plots and data are successfully produced, stored in the local disk, and freely distributed via the SWELTO portal;
	\item \textbf{SIDmonitor} module status: \textbf{Level2.5}. The pipeline to read and provide in output plots and data is ready, but the antenna and SIDmonitor have still be put back into operation after the stop due to the ongoing pandemic emergency. The production of a new radio antenna with a 3D printer is still in progress; 
	\item \textbf{FGmagnetometer} module status: \textbf{Level1}. Calibration and testing activities of the sensor and ADC are still in progress after the stop due to the ongoing pandemic emergency. The pipelines to extract and analyze the data, and provide analysis results have still to be conceived and written.
	\item \textbf{SkyMonitor} module status: \textbf{Level1}. The concept has been successfully tested, but pipelines to automatically identify the atmospheric TLEs have still to be conceived and written.
	\item \textbf{CosmicRay} module status: \textbf{Level0.5}. The possible implementation of cosmic ray data sharing in real-time have still to be explored; all the pipelines for automated data analysis have to be written.
	\item \textbf{RadioBurst} module status: \textbf{Level0}. The aim of this module (still to be conceived) will be to analyze radio dynamic spectra to automatically identify and track solar Type-II radio bursts, associated with propagating shock waves.
	\item \textbf{EUVCorDens} module status: \textbf{Level0}. The aim of this module (still to be conceived) will be to analyze EUV images of the full Sun to provide automatically the distribution of electron density and plasma emission measure in the inner corona by applying a Differential Emission Measure technique.
\end{itemize}
For what concerns the SWELTO sensors network (SIDmonitor, FGmagnetometer, SkyMonitor, and CosmicRays), the activities have been stopped or delayed due to the on-going pandemic emergency. As soon as the access to the Institute will be allowed again with continuity, the calibrations and installations of the hardware already procured will restart and be completed.

Dissemination of the Project is now in progress. The project has been already presented at the last \href{http://www.stce.be/esww2019/program/session_details.php?nr=17}{"16th European Space Weather Week"} conference (with a poster presentation), and will be presented on-line at the next \href{http://esws2020.iopconfs.org/quickviews}{"European Space Weather Symposium 2020"} (with a Quick View presentation). In the next future the SWELTO Team will be registered on the \href{https://ccmc.gsfc.nasa.gov/iswat/}{ISWAT Portal} (International Space Weather Action Teams). Other future actions will be considered to give an increasing visibility to the Project. Moreover, future national and international calls supporting Space Weather activities will be considered for future funding.

\phantomsection
\section*{Acknowledgments} 

\addcontentsline{toc}{section}{Acknowledgments} 

The SWELTO project is at present entirely supported by the INAF-Turin Astrophysical observatory. The SuperSID monitor was provided as in kind contribution by University of Stanford, whose support is gratefully acknowledged.


\phantomsection
\bibliographystyle{unsrt}
\bibliography{references.bib}


\end{document}